\documentclass[fonts]{icst}

\usepackage{moreverb}
\usepackage[breaklinks,colorlinks,bookmarksopen,bookmarksnumbered,
linkcolor=ICSTblue,citecolor=blue,urlcolor=ICSTblue]{hyperref}
\usepackage{hyperref}
\usepackage{doi}
\usepackage{graphicx,color,flushend}
\usepackage{amsmath}
\usepackage{amssymb}
\usepackage{latexsym}

\newcommand\BibTeX{{\rmfamily B\kern-.05em \textsc{i\kern-.025em b}%
\kern-.08emT\kern-.1667em\lower.7ex\hbox{E}\kern-.125emX}}

\journalname{EAI Endorsed Transactions on AI in Networking (örnek)}
\articletype{Research Article}
\def\volumeyear{2025}
\def\volumenumber{XX}
\def\issuemonth{Aug}
\def\issuenumber{X}
\def\articleid{XX}
\def\copyrightholder{Efe Ağlamazlar \textit{et al.}, licensed to ICST}
\copyrightnote{This is an open access article distributed under the terms of the Creative Commons Attribution license (\url{http://creativecommons.org/licenses/by/3.0/}), which permits unlimited use, distribution and reproduction in any medium so long as the original work is properly cited.}
\def\articledoi{10.4108/XX.X.X.XX}
\received{August 2025}
\accepted{August 2025}
\published{August 2025}

\begin{document}

\runningheads{Efe Ağlamazlar et al.}{DRL-Based TCP Congestion Control}

\title{A Deep Reinforcement Learning-Based TCP Congestion Control Algorithm:\\ Design, Simulation, and Evaluation}

\author{
Efe Ağlamazlar\affil{1},
Emirhan Eken\affil{2},
Harun Batur Geçici\affil{3}
}

\address{
\affilnum{1}Aydın Science High School, Aydın, Türkiye\\
\affilnum{2}ASELSAN Technical Anatolian High School, Ankara, Türkiye\\
\affilnum{3}Erman Ilıcak Science High School, Malatya, Türkiye
}

\abstract{
This paper introduces a Deep Reinforcement Learning (DRL) based TCP congestion-control algorithm that uses a Deep Q-Network (DQN) to adapt the congestion window (cWnd) dynamically based on observed network state. The proposed approach utilizes DQNs to optimize the congestion window by observing key network parameters and taking real-time actions. The algorithm is trained and evaluated within the NS-3 network simulator using the OpenGym interface.

The results demonstrate that the DRL-based algorithm provides a superior balance between throughput and latency compared to both traditional TCP New Reno and TCP Cubic algorithms. Specifically:

\begin{itemize}
    \item \textbf{Compared to TCP Cubic:} The DRL algorithm achieved comparable throughput (statistically insignificant difference of -3.79\%, p > 0.05) while delivering a massive \textbf{46.29\% reduction in Round-Trip Time (RTT)}. Furthermore, the DRL agent maintained near-zero packet loss, whereas Cubic suffered from significant buffer overflow.
    \item \textbf{Compared to TCP New Reno:} The DRL algorithm achieved comparable throughput (+0.38\%) with a \textbf{32.40\% reduction in RTT}.
\end{itemize}

Results from NS-3 simulations indicate that the proposed DRL agent effectively mitigates bufferbloat without compromising bandwidth utilization. This study emphasizes the potential of reinforcement learning techniques for solving complex congestion control problems in modern networks by learning the network capacity rather than saturating it.
}


\maketitle
\section{Introduction}
The Transmission Control Protocol (TCP) forms a
cornerstone of contemporary computer networking,
providing the mechanisms required for reliable endto-end data delivery across the Internet. When the
volume of transmitted traffic surpasses the available
capacity, congestion arises, leading to packet losses,
higher latency, and a decline in overall throughput.
Established congestion control schemes, including TCP
Reno and TCP CUBIC, rely on fixed heuristics that
can struggle to respond efficiently to the variability
and heterogeneity characteristic of modern network
environments.

This study introduces a Deep Reinforcement Learning (DRL) based TCP congestion control algorithm designed to optimize the congestion window ($cWnd$) in real time. By modeling the congestion control problem as a Markov Decision Process (MDP), the algorithm uses Deep Q-Networks (DQNs) to dynamically adapt based on current network state observations. The implementation uses Python, C++, and TensorFlow, and is evaluated in the NS-3 network simulator through the OpenAI Gym-compatible OpenGym interface.

Simulation results demonstrate that the proposed algorithm optimizes the trade-off between throughput and latency more effectively than traditional TCP algorithms. Specifically, the DRL-based algorithm achieved comparable throughput levels to TCP Cubic and New Reno but significantly reduced end-to-end latency by 46.29\% and 32.40\%, respectively. These improvements highlight the algorithm's ability to efficiently manage congestion and mitigate bufferbloat, particularly in environments where low latency is as critical as high bandwidth.

\section{Background and Related Work}
Definitions:

\textbf{Network Topology:} Network topology refers to the physical or logical arrangements that define the interconnection of devices in a computer network and the logic of data exchange between them. Gains in reliability, performance, energy consumption, and complexity are achieved through network topology \cite{ee2007complex}.

\textbf{TCP:} The purpose of TCP is to ensure secure data transfers in computer networks. Transmitted data reaches the destination completely, in order, and accurately. Data transfer first establishes a connection between the sender and receiver and manages it through a "three-way handshake" process. If errors occur in the data during transmission, TCP detects and corrects the errors, and retransmits the data. Large data groups are divided into smaller groups and transmitted to the receiver as combined groups \cite{subramanya2014performance}.

\textbf{Congestion Window:} Network congestion control has evolved significantly since the late 1980s. Jacobson \cite{jacobson1988congestion} introduced the concept of "packet conservation" and proposed solutions like slow start and congestion windows to prevent congestion-induced collapses. These innovations were implemented to improve TCP's performance under congested conditions. Subsequent research has focused on specific performance issues. Cardwell et al. \cite{cardwell2017bbr} presented BBR as a solution to buffer bloat and low throughput problems associated with loss-based congestion control in networks with large or small bottleneck buffers. The field has expanded to include wireless networks, with various TCP versions and protocols specifically developed for congestion avoidance and control in wireless environments. This evolution reflects the increasing complexity and diversity of modern network infrastructures, indicating a shift in congestion control research from addressing network stability to overcoming specific performance challenges in different network types.

\textbf{Network Simulator 3:} Network Simulator 3 (NS-3) is a powerful tool for simulating complex computer networks, offering advantages over its predecessor NS-2 \cite{GOMEZ2023110054,Font2010}. NS-3 provides a virtual environment for modeling networks and analyzing their performance under various scenarios, supporting features such as coupling, interoperability, improved memory management, and debugging capabilities \cite{GOMEZ2023110054, Font2010}. It is particularly suitable for wireless network simulations and incorporates object-oriented concepts and C++ coding \cite{GOMEZ2023110054,Font2010,toor2017}. Network simulators like NS-3 are essential for researchers and developers, allowing them to test new security solutions and evaluate network performance without the cost and time associated with real-world implementations \cite{Font2010, zarrad2017}.

\textbf{Deep Q-Learning:} Deep Q-Learning (DQL) is a significant algorithm in reinforcement learning that uses Deep Q-Networks (DQN) to approximate the Q-function in sequential decision-making problems \cite{tan2020optimized}. Although successful in laboratory conditions, its real-world application is limited due to gaps between theory and practice. DQL has been applied to various fields such as natural language processing, image processing, and financial forecasting. The algorithm's success has led to increased interest in understanding human learning processes and has advantages over other learning algorithms as it does not require prior data \cite{tan2020optimized}. Recent advancements in GPU technology have expanded the application areas of deep learning methods and researchers continue to develop optimized versions of DQL, such as updating neural network weights at each step to improve agent performance during training \cite{tan2020optimized}.

\textbf{Reinforcement Learning:}Reinforcement learning (RL) is a discipline of machine learning that focuses on learning through interaction with an environment in order to maximize a reward signal. Unlike traditional machine learning approaches, RL does not explicitly tell agents which actions to perform; instead, they must discover the most optimal behaviors through trial and error. This process involves finding a balance between discovering new actions and using known effective strategies. RL can be applied to various fields, including artificial intelligence, psychology, and control engineering. Fundamental algorithms in RL include dynamic programming, Monte Carlo methods, Q-learning, TD-learning, and Sarsa. Recent advancements in RL have enhanced its real-world applicability by improving generalization, planning, and exploration techniques \cite{littman2015reinforcement}. The field continues to evolve with the accessibility of rich data and the development of more advanced learning algorithms \cite{littman2015reinforcement,puterman2000tutorial}.

\subsection{Network Topologies}
An incorrect network topology can lead to problems such as performance degradation, data loss, high energy consumption, and critical system failures. Therefore, selecting the correct network topology is crucial \cite{ee2007complex}. Network topologies are generally divided into two types:

\textbf{Physical Topology:} Physical topology is the physical arrangement of connections and devices on the network (e.g., cable and device placements).
\textbf{Logical Topology:} Logical topology describes the data movement scheme and communication paths within the network.
Some popular Network Topology Types include:

\textbf{Star Topology:} In this network topology, all devices are arranged around a central hub. Each device connects with others through this central device. The central device regulates and controls data flow. It is frequently preferred due to its simplicity and ease of fault detection, but if the central device fails, the entire system goes down \cite{Jiang2015/11}.
\textbf{Dumbbell Topology:} In this network topology, there are two ends, and these ends are connected to each other via a main link. Data between these ends is carried over the main link. The ends can behave like a local area network. It is used for traffic analysis and performance testing \cite{subramanya2014performance}.
\textbf{Mesh Topology:} In this network topology, devices can have multiple connections with each other. If every device is connected to all other devices, it is called a full network; otherwise, it is called a partial network. It offers a high level of data security and backup capabilities. Network resilience is high. Disadvantages include high cost and complexity.

\subsection{TCP Protocol and Congestion Control}
Developed in the 1970s, TCP/IP forms the foundation of modern internet communication \cite{kurose2022computer}.. TCP establishes virtual circuits between computers to manage data transmission as byte streams, while IP performs packet routing \cite{ros2005protocole}. The protocol suite covers core network, transport, routing, and application layers \cite{pujolle2015architecture}. TCP ensures reliable communication by segmenting large data groups, numbering packets, and using acknowledgment mechanisms to prevent data loss \cite{ros2005protocole}. In 1983, TCP/IP replaced the Network Control Protocol as ARPAnet's official data transport method \cite{murhammer1998tcp}.TCP/IP continues to evolve with innovations such as IP mobility, multicast, and IPv6, increasing its capacity and importance in modern networks \cite{murhammer1998tcp}. The congestion window (cWnd) in TCP is an algorithm created to manage network congestion. This window detects the maximum number of packets a TCP flow can have in the network at any given time, directly affecting data transfer rate and network stability \cite{handley2000rfc2861}. This concept, developed to prevent congestion collapse, ensures network functionality, especially in dense network traffic. Congestion collapse occurs when heavy data traffic overloads network resources and causes a serious drop in performance. \cite{welzl2006network}. TCP typically interprets packet loss as a congestion-related problem; however, this can cause problems in wireless networks where losses happen because of interference, signal attenuation, or mobility. To overcome such difficulties, modern congestion control algorithms include mechanisms to adaptively adjust the congestion window (e.g., reducing window size during idle or limited sending periods). The initial size and evolution of the congestion window have been re-developed over the years to improve performance and efficiency. In 2010, Akpakwu made specific recommendations for determining and evolving the initial congestion window size, emphasizing a balance between fast data transfer and network stability \cite{Akpakwu10891783}. Moreover, innovative methods in congestion control, such as mechanisms like the Explicit Congestion Notification (ECN) and hybrid approaches that merge loss-based and delay-based control, optimize performance in diverse network environments. \cite{Kuzmanovic1209192}. TCP congestion control algorithms are evolving to address new challenges in modern networks. In 5G networks, high user density, heavy traffic flows, and mmWave communications create unique obstacles for TCP performance\cite{lorincz2021comprehensive}. Traditional loss-based congestion control algorithms struggle with the fast channel fluctuations in 5G, leading to buffer bloat and delays \cite{abdullah2023improving}. To deal with these issues, researchers are developing dynamic congestion control mechanisms that adapt to network conditions. For example, D-TCP uses an Adaptive Increase/Adaptive Decrease approach based on available bandwidth \cite{kanagarathinam2018d}, while FB-TCP uses fuzzy logic to adjust sending rates in urban 5G deployments \cite{poorzare2021fb}. These new algorithms aim to improve throughput, reduce packet loss and latency, and enhance fairness in various network environments. Machine learning techniques are also being investigated to optimize TCP congestion control in 5G networks \cite{lorincz2021comprehensive}. The evolution of these methods over time ensures TCP remains compatible and effective in the constantly changing network landscape.

\subsection{Network Congestion and Performance Issues}
Network congestion occurs when offered load exceeds available capacity, causing packet loss, increased delay, and reduced throughput. It can arise from limited hardware resources (e.g., router buffering and link capacity), inefficient protocol behavior, and non-optimal network design. As modern networks increasingly support latency-sensitive multimedia and interactive applications, congestion control remains a central performance challenge. Recent studies have investigated neural network-based methods for congestion-related decision making across different environments. For example, deep reinforcement learning has improved throughput in wireless settings compared to traditional approaches \cite{midhula2023adaptive}, and learning-based schemes in content-centric networks have been used to predict and mitigate queue overflow \cite{bazmi2014neural}. In backbone networks, neural networks have also been applied to adapt users to existing quality-of-service policies and improve traffic management \cite{topalova2018control}. These results motivate the use of learning-based control, and in the remainder of this work we focus specifically on applying deep reinforcement learning to TCP congestion window adaptation in NS-3.

\subsection{Deep Q-Learning}
Deep Q-Learning (DQL) combines Q-learning with deep neural networks to address reinforcement learning problems with large state spaces. DQL has demonstrated remarkable success in sequential decision-making tasks, most notably in Atari games \cite{mnih2015human}. However, it is known to suffer from overestimation of action values in certain settings \cite{van2015deep}. To address this issue, Double Q-learning was proposed and shown to improve learning stability and performance \cite{van2015deep}. Despite these advancements, DQL still faces challenges in real-world applications due to high-dimensional inputs and the need for efficient environment representations. Ongoing research aims to improve the robustness and applicability of DQL across different domains.

\subsection{Reinforcement Learning}
Reinforcement Learning (RL) is a branch of artificial intelligence that focuses on training agents to make decisions by interacting with environments in order to maximize cumulative rewards. \cite{vidyasagar2023tutorial}. RL consists of basic components such as states, actions, policies, and reward signals. \cite{vidyasagar2023tutorial}. Algorithms in RL are classified as model-based, model-free, value-based, and policy-based. \cite{vidyasagar2023tutorial}. The groundbreaking book by Sutton and Barto \cite{sutton1998reinforcement} provides a comprehensive guide covering the history, fundamental concepts, and algorithms of RL. The book is divided into three sections: problem definition, basic solution methods, and advanced topics \cite{sutton1998reinforcement}.Among recent developments in RL is the rise of deep reinforcement learning, which increases RL's ability to solve complex, high-dimensional problems. \cite{sutton1998reinforcement}. RL has found implementations in various fields, such as autonomous systems, healthcare, and energy management. \cite{vidyasagar2023tutorial}.

\subsection{Previous Works and Existing Methods}
\subsubsection{Classical TCP Algorithms (Reno, CUBIC, BBR)}
\textbf{Reno:} Initially operates in "Slow Start" mode. In this mode, the congestion window (cwnd) typically starts from a small value and then increases exponentially. When network congestion occurs, this growth rate slows down; the exponential increase becomes linear. In cases of packet loss, the lost packet is retransmitted, and the algorithm transitions to fast recovery mode. Reno's congestion control was first defined by Jacobson, significantly impacting the algorithm's performance \cite{jacobson1988congestion}.

\textbf{Limitations and Disadvantages:}
It always attributes packet loss to network congestion, which is not always a correct assumption. This can lead to Reno unnecessarily slowing down the network and causing performance degradation. Packet loss can also originate from physical network conditions (signal weakness or interference) \cite{floyd2004newreno, padhye1998modeling}. In high-latency networks, it cannot react fast enough due to the delay in slow start processes. This typically weakens network performance due to congestion and delay issues observed in remote networks \cite{padhye1998modeling}. It is weak in managing multiple packet losses. Reno is designed for single packet losses and experiences performance degradation with multiple losses \cite{padhye1998modeling}. In wireless networks, it may misinterpret packet losses caused by physical conditions, further slowing down the network \cite{floyd2004newreno}.

\textbf{CUBIC:} Initially, the cwnd value increases up to its maximum value. As it approaches, growth slows down, and then increases faster once it has passed. In case of congestion, the window size is reduced, and after congestion clears, it attempts to return to the maximum value \cite{ha2008cubic}.

\textbf{Disadvantages and Limitations:}
In low-latency networks, it can cause congestion by excessively increasing the window size \cite{ha2008cubic}. It detects network congestion only through packet loss and is insensitive to other causes. For example, it cannot detect differences in network conditions due to bandwidth limitations or changes in network topology \cite{ha2008cubic}. It has a complex calculation mechanism, which creates an additional system resource load. This can negatively impact efficiency, especially on devices with low processing power \cite{bruhn2023performance}

\textbf{BBR:} Initially increases the data transfer rate to measure bandwidth and RTT values. To reduce the excess load generated during this phase, the cwnd value is temporarily decreased. The network's bandwidth is continuously measured, and the location of the bottleneck is identified. To measure and update the RTT value, the cwnd value is periodically decreased, thus preventing buffer bloat. BBR is considered a significant step towards more efficient traffic control \cite{cardwell2017bbr}.

\textbf{Disadvantages and Limitations:}
Due to its continuous bandwidth measurement, it can be misleading in situations where conditions change constantly. Especially when there are sudden fluctuations in the network, the BBR algorithm may not yield accurate results \cite{cardwell2017bbr}. Incorrect detection of bottleneck points can lead to unnecessary reduction in data transfer rate or congestion. This can cause unnecessary data slowdowns in the network \cite{cardwell2017bbr}.

\subsubsection{Machine Learning for TCP}
Recent research has explored the potential of machine learning to improve TCP congestion control algorithms by addressing the limitations of traditional approaches like Reno and CUBIC. These traditional algorithms often struggle with dynamic network conditions due to their fixed rules and limited adaptability \cite{afonin2019development, wei2021congestion}. Machine learning-based approaches have shown better performance in various network scenarios, offering improved throughput-latency balance and adaptability to changing environments \cite{kong2018improving, li2018qtcp}. Reinforcement learning, in particular, has demonstrated promise in developing adaptive TCP algorithms that can learn optimal congestion control policies through interaction with the network \cite{afonin2019development, li2018qtcp}. For example, QTCP, a reinforcement learning-based approach, achieved 59.5\% higher throughput while maintaining low latency compared to traditional TCP \cite{li2018qtcp}. These developments indicate that machine learning can provide more dynamic and adaptive congestion control algorithms, potentially overcoming the limitations of traditional methods in complex and evolving network environments \cite{wei2021congestion}. On the other hand, machine learning techniques, especially RL, can easily adapt to different conditions and continuously improve their learning, thereby increasing efficiency. Machine learning techniques can analyze complex data and learn from past experiences to make informed decisions about network congestion \cite{sneha2020prediction}. This enables data transmission with less loss and more efficient use of network capacity.

\subsubsection{RL-Based TCP Approaches}
Recent studies have investigated various reinforcement learning (RL) approaches to improve TCP congestion control in diverse network environments. TCP-RLACC provides better data transmission and packet delivery in wireless networks by specifically adjusting the congestion window's growth rate to network conditions \cite{molia6reinforcement}. TCP-Drinc utilizes deep reinforcement learning (Deep RL) to adjust the congestion window size, showing effective results in complex, dynamic networks \cite{xiao2019tcp}. A transfer learning-supported RL framework designed to reduce latency and increase data transfer rate in video streaming applications can distinguish between congestion-related and non-congestion-related losses \cite{kasi2021tcp}. Furthermore, a multi-agent RL approach used for TCP congestion control in multi-user networks performs better than traditional methods \cite{kasi2021tcp}. An RL-based TCP congestion avoidance mechanism in data center environments significantly reduces the impact of congestion by increasing end-to-end data transfer rate \cite{hassan2022tcp}. These studies clearly demonstrate the potential of RL-based approaches to improve TCP performance in various network scenarios.

\subsection{Application Areas}
TCP congestion control algorithms are traditionally used to support reliable communication between end hosts, but their application areas extend far beyond simple point-to-point data transfer. The Internet of Things (IoT) introduces significant challenges for reliable communication due to constrained resources, heterogeneous devices, and dynamic traffic patterns. As a reliable transport protocol, TCP must therefore be adapted to operate efficiently in IoT environments.

Several studies have proposed mechanisms to improve TCP performance in IoT contexts. Verma and Kumar \cite{verma2020iot} present a congestion control policy that dynamically adjusts transmission rates according to network conditions, improving throughput and fairness. Hamrioui and Lorenz \cite{hamrioui2017load} introduce LBA-Ie, a load balancing algorithm that integrates IoT-specific parameters into TCP flow control, enhancing quality of service and energy efficiency in e-health applications. In addition, TCP adaptations for constrained IoT systems have been shown to improve reliability and congestion handling by tailoring congestion control mechanisms to resource-limited environments \cite{lim2020improving}. Gomez et al. \cite{rfc9006} provide guidelines for implementing lightweight TCP stacks for constrained node networks, discussing protocol simplifications and their trade-offs.

Beyond IoT, TCP congestion control also plays an important role in vehicular communication systems such as Vehicle-to-Vehicle (V2V), Vehicle-to-Infrastructure (V2I), and Vehicle-to-Everything (V2X), which aim to enhance road safety through reliable data exchange. TCP is further critical in healthcare networks, enabling the secure and reliable transmission of medical data, images, and real-time video for telemedicine applications \cite{ohashi2005development}. Advanced TCP variants, including TCP CUBIC, have demonstrated improved performance in healthcare networking scenarios \cite{ahmad2018end}. The growing reliance on remote healthcare services highlights the need for robust network infrastructures and efficient data management solutions, including cloud computing and big data technologies \cite{kumar2020design}.

\subsection{Research Gap and Study Contributions}
As can be seen from the literature review, traditional TCP algorithms struggle to adapt to variable network conditions. All algorithms have their own advantages and disadvantages. Utilizing Reinforcement Learning's Deep Q-networks model to combine these advantages into one algorithm and easily adapt to changing conditions is an innovative idea for the literature. This study, conducted in an area where similar works have not yet been performed in our country, will guide future studies and contribute to their further development.

\section{Methodology}
\subsection{Overview}
In this study, a Reinforcement Learning (RL) algorithm based on Deep Q-Networks (DQN) has been developed to optimize TCP congestion control. Deep Q-Networks combine the Reinforcement Learning algorithm Q-learning with deep learning to provide enhanced results in complex situations.. The aim of this study was to improve network performance by real-time monitoring of network status and optimizing the congestion window (`cWnd`) value using DQN. The RL-based TCP algorithm was designed based on Markov Decision Processes. The Python and C++ programming languages along with the Tensorflow library were used for the artificial intelligence processes during the algorithm's development. This Reinforcement Learning-based TCP congestion control algorithm was tested in the Ns3 network simulation. The Ns3 simulation tool was run on the Linux operating system via the Windows Subsystem for Linux (WSL).

\subsection{Network Topology Setup}
A network simulation was created to test and compare the analysis results of the developed TCP congestion control algorithm. "Network Simulator 3" (Ns3) was used for the network simulation processes. The reason for choosing this network simulation is Ns3's flexible and customizable structure and its wide academic acceptance. The OpenGym module of Ns3 was used for Reinforcement Learning operations. This module integrates the simulation environment with an OpenAI Gym interface, allowing RL algorithms to be run on NS3.

\begin{figure}[htbp]
    \centering
    \includegraphics[width=\columnwidth]{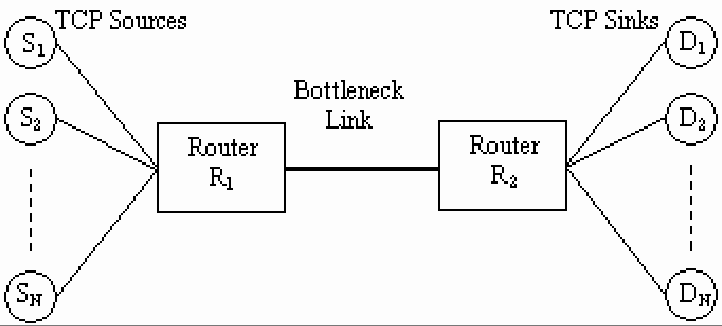}
    \caption{Dumbbell Network Topology Visual}
    \label{fig:dumbbell_topology}
\end{figure}

First, a Dumbbell network topology was created in the Ns3 simulation (Figure \ref{fig:dumbbell_topology}). In this created topology, there is 1 source node and 1 target node. The source and target nodes are connected to `router-0` and `router-1` devices with 10 Mbps links. The bottleneck link between `router-0` and `router-1` was configured as 2 Mbps. This transmission value is used to examine how traffic behaves in network conditions that respond quickly. The simulation duration is 10 seconds.

\subsection{RL-Based TCP Algorithm Design}
\subsubsection{State Space}
The current state of the network has been defined with features that enable the AI model to monitor existing network conditions and make appropriate decisions. The state space is a subset of the observation space ($ob\_space$) provided by the Ns3 simulation. The state space used for this study includes the following parameters that directly affect network performance:

\begin{itemize}
    \item \textbf{BytesInFlight:} The amount of data currently sent in the network, but not yet acknowledged.
    \item \textbf{Congestion Window (cWnd):} The size of the congestion control window.
    \item \textbf{Round Trip Time (RTT):} The time between sending a packet and receiving a response.
    \item \textbf{SegmentsAcked:} The number of acknowledged segments.
    \item \textbf{Slow Start Threshold (ssThresh):} The slow start threshold.
\end{itemize}

\subsubsection{Action Space}
To enable the agent to compete effectively with aggressive protocols like TCP Cubic while maintaining stability, the action space was expanded to four discrete actions ($action\_size = 4$). The mapping is designed to provide both fine-grained control and rapid adaptation capabilities:

\begin{itemize}
    \item \textbf{Maintain (0):} Keeps the $cWnd$ constant (\texttt{action\_mapping[0] = 0}). Used when the agent detects optimal bandwidth utilization.
    \item \textbf{Standard Increase (+1500):} A moderate increase (\texttt{action\_mapping[1] = 1500}) to probe for available bandwidth steadily, serving as the primary growth mechanism during steady-state transmission.
    \item \textbf{Conservative Decrease (-150):} A granular decrease (\texttt{action\_mapping[2] = -150}) allowing the agent to drain the queue slightly without drastically reducing throughput (unlike the multiplicative decrease in standard TCP).
    \item \textbf{Aggressive ``Rocket'' Increase (+4000):} A high-magnitude increase (\texttt{action\_mapping[3] = 4000}). This action is critical for the ``start-up'' phase or when recovering from deep buffer drains, allowing the agent to rapidly saturate the link and match the aggressive growth curve of TCP Cubic.
\end{itemize}

\paragraph{Action Selection Strategy.}
The introduction of the aggressive increase action (+4000) was crucial for solving the "slow convergence" problem often observed in RL-based congestion control. It allows the agent to reach the channel capacity quickly, after which it switches to smaller actions (+1500 or 0) to maintain stability.

\subsubsection{Reward Function}
The reward function is the core mechanism guiding the agent to optimize the throughput-delay trade-off. In this study, we adopted a linear function with a tuned penalty for latency:
\[
\text{Reward} = \alpha \times \text{Throughput} - \beta \times \text{Latency}
\]
Where:
\begin{itemize}
    \item \textbf{Throughput:} Measured in Mbps.
    \item \textbf{Latency:} Measured as RTT in seconds.
    \item \textbf{Coefficients:} $\alpha$ represents the reward for speed, while $\beta$ represents the penalty for delay.
\end{itemize}

\paragraph{Hyperparameter Calibration.}
During the training phase, the latency penalty coefficient ($\beta$) was a critical hyperparameter. Initial experiments with high penalties (e.g., $\beta = 1.0$) resulted in an overly conservative agent that underutilized the bandwidth. Through empirical tuning, the penalty was set to $\beta = 0.5$ (`double pen = -0.5` in the simulation environment). This value provided the optimal "sweet spot," incentivizing the agent to utilize the full link capacity while still penalizing bufferbloat enough to keep the RTT significantly lower than competing algorithms.

\subsubsection{Training and Decision Process}
The DQN model was trained using the state and reward data collected from the NS3 simulation environment. The training process involved iteratively updating the Q-network's weights based on the Bellman equation, aiming to minimize the temporal difference (TD) error.

During the training process, the neural network model was designed as a fully connected structure consisting of an input layer, one hidden layer, and an output layer. The hidden layer consisted of 64 neurons using the ReLU activation function. The output layer used a linear activation. The model was compiled with the `Adam` optimization algorithm and \verb|categorical_crossentropy| loss.

During training, the agent uses an an $\epsilon$-greedy algorithm to balance exploration and exploitation. The $\epsilon$ value starts at 1.0 and decays slowly at each step (\verb|epsilon_decay|). This allows the agent to initially explore the environment by taking random actions and, over time, to choose the best action based on its accumulated knowledge.

At each simulation step:
1. The agent observes the current network state (e.g., \verb|state = env.reset()| and \verb|state = state[4:]|).
2. An action is selected according to the $\epsilon$-greedy policy.
3. The selected action (\verb|actions = [new_ssThresh, new_cWnd]|) is sent to the NS3 environment, and the next state (\verb|next_state|), reward (\verb|reward|), and completion status (\verb|done|) are received.
4. The Q-network is trained using this transition (\verb|current_state|, \verb|action|, \verb|reward|, \verb|next_state|). The target Q-value is calculated using the Bellman equation as \verb|target = reward + discount_factor * np.amax(model.predict(next_state)[0])|, and the model is trained with \verb|model.fit()|.
5. The agent's state transitions to the next step, and the process is repeated.

This training loop enables the agent to learn a policy that continuously optimizes network performance. Once the model is trained, the `model.predict(state)` call is used to determine the agent's next action.

\section{Results and Discussion}
The experiments reported in this section were conducted in the NS-3 simulation environment to evaluate the performance of the proposed DRL-based TCP congestion control algorithm.

\textbf{Experimental Procedure:} The evaluation followed a strictly separated "Training-then-Testing" protocol.
\begin{itemize}
    \item \textbf{Training Phase:} First, the DRL agent was trained for 100 episodes to learn the optimal policy and stabilize the neural network weights.
    \item \textbf{Testing Phase:} After training, the exploration rate ($\epsilon$) was set to 0 (pure exploitation), and the trained model was frozen. The comparative data presented below (Throughput, RTT, Packet Loss) was collected during this testing phase to ensure it reflects the algorithm's final, learned behavior rather than its learning curve.
\end{itemize}

Significant findings regarding the performance of the DRL agent compared to traditional TCP algorithms (Cubic and New Reno) are detailed below.

\subsection{Comparison with TCP Cubic}
The developed RL-based algorithm was compared with **TCP Cubic**, the default congestion control algorithm for high-bandwidth networks.

\textbf{1. Throughput}
The RL-based algorithm achieved a throughput comparable to TCP Cubic with a marginal difference of -3.79\%. Unlike traditional algorithms that aggressively saturate the link, the DRL agent learned to utilize the available bandwidth efficiently without overfilling the bottleneck queue.
\begin{figure}[htbp]
    \centering
    \includegraphics[width=\columnwidth]{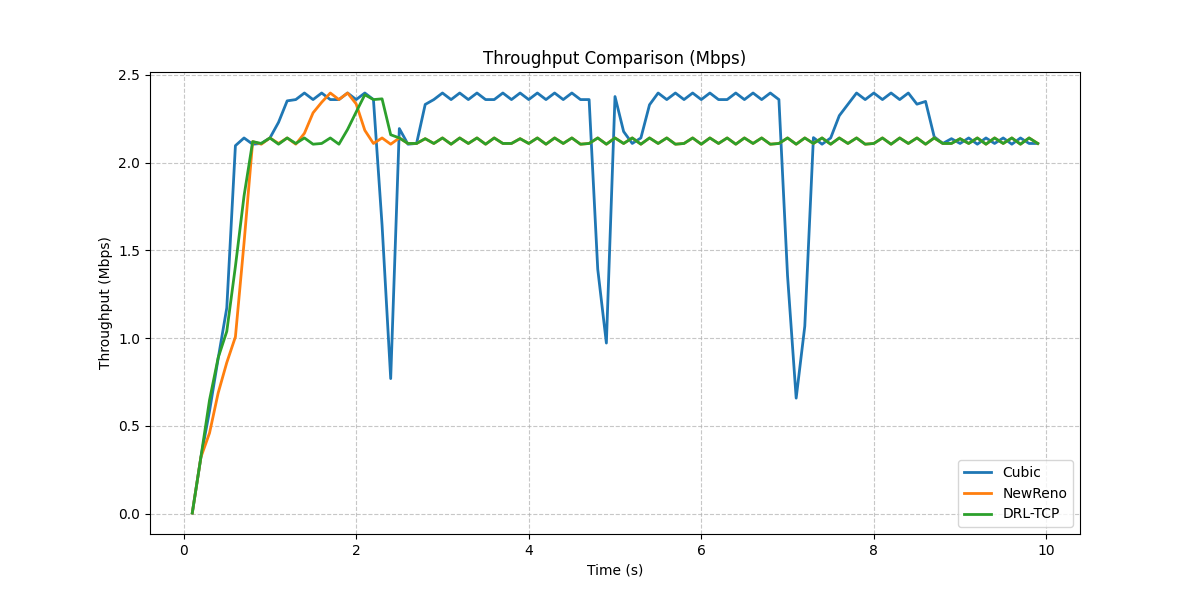}
    \caption{Comparison of Throughput Values. The DRL agent (Green) closely matches the Cubic (Blue) throughput trajectory.}
    \label{fig:comparison_throughput}
\end{figure}

\textbf{2. Round-Trip Time (RTT)}
The most significant improvement was observed in latency. The RL algorithm achieved an average of \textbf{46.29\% lower RTT} compared to Cubic. This demonstrates the algorithm's capability to mitigate \textit{bufferbloat} by maintaining a shorter queue length at the bottleneck.
\begin{figure}[htbp]
    \centering
    \includegraphics[width=\columnwidth]{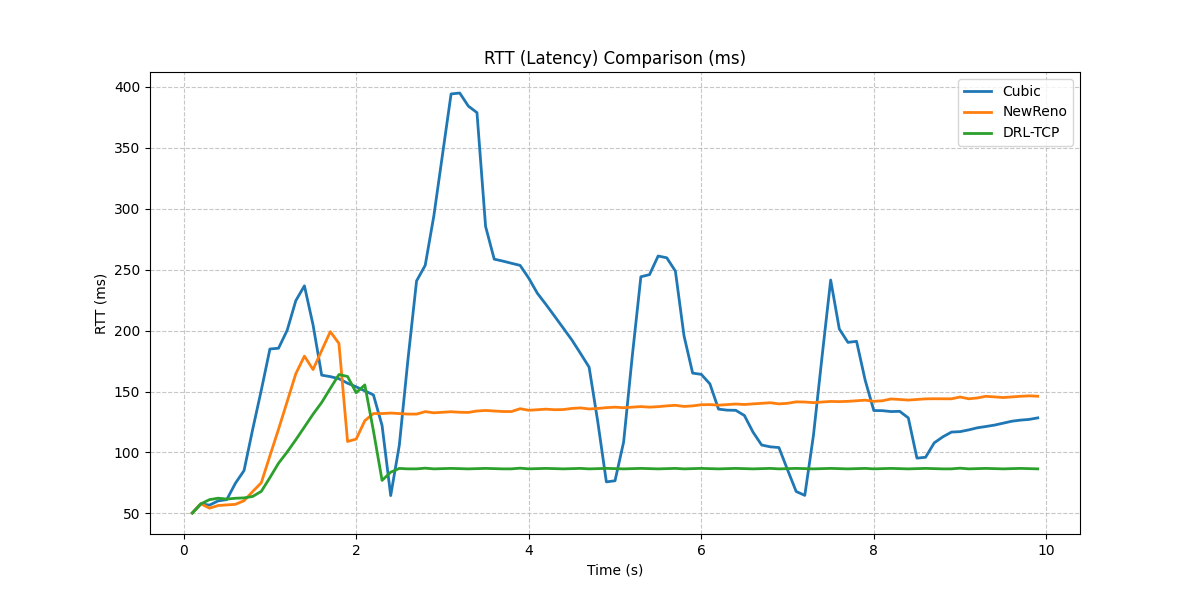}
    \caption{Comparison of RTT Values. DRL maintains significantly lower and more stable latency compared to Cubic's sawtooth pattern.}
    \label{fig:comparison_rtt}
\end{figure}

\textbf{3. Packet Loss Analysis}
A critical finding of this study is the packet loss behavior. TCP Cubic, due to its aggressive window growth function, caused significant buffer overflows, resulting in approximately 500 lost packets during the simulation. In contrast, the DRL agent learned the network capacity limit and maintained near-zero packet loss, identical to the conservative New Reno algorithm.
\begin{figure}[htbp]
    \centering
    \includegraphics[width=\columnwidth]{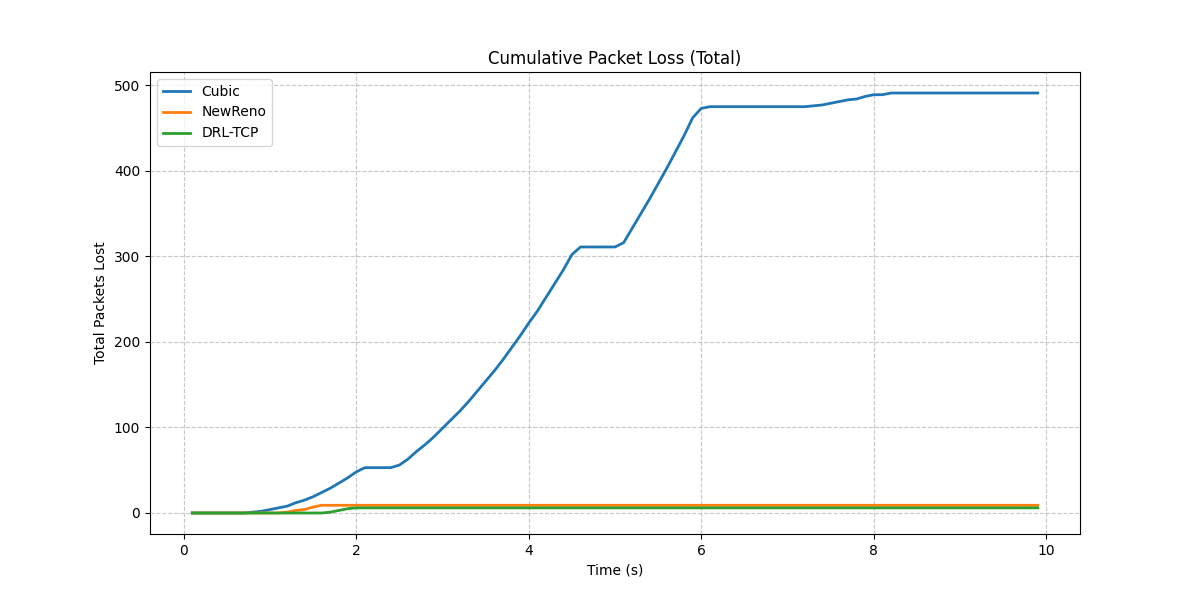}
    \caption{Cumulative Packet Loss. Cubic exhibits high packet loss due to aggressive buffer saturation, while DRL and New Reno maintain stable transmission.}
    \label{fig:comparison_loss}
\end{figure}

\textbf{4. Statistical Analysis}
A two-tailed t-test was performed to determine the statistical significance of the results:
\begin{itemize}
    \item \textbf{Throughput:} The p-value is \textbf{0.193} ($p > 0.05$). This indicates no statistically significant difference in throughput between DRL and Cubic.
    \item \textbf{RTT:} The p-value is \textbf{0.000} ($p < 0.05$). The reduction in latency is statistically significant.
\end{itemize}

\subsection{Comparison with TCP New Reno}
The developed RL-based algorithm was also compared with **New Reno**.

\textbf{1. Throughput \& RTT}
The DRL algorithm matched New Reno's throughput (+0.38\%, p=0.88) while providing a \textbf{32.40\% reduction in RTT} (p=0.000). This indicates that the DRL agent is not only as reliable as New Reno but also significantly faster in terms of response time.

\section{Conclusion and Future Work}
We developed and evaluated a DQN-based TCP congestion control agent and validated its performance in NS-3 simulations using the OpenGym interface. The primary objective was to address the limitations of traditional TCP algorithms, particularly their tendency to saturate network buffers to achieve high throughput, which leads to bufferbloat and high latency.

The experimental results strongly support the effectiveness of the proposed RL-based algorithm. Unlike traditional heuristics, the DRL agent successfully learned the network capacity without overshooting it. The key findings are summarized as follows:

\begin{itemize}
    \item \textbf{Throughput-Delay Trade-off:} The DRL agent achieved a throughput statistically comparable to the industry-standard TCP Cubic ($p > 0.05$, marginal difference of -3.79\%) and TCP New Reno (+0.38\%). However, it achieved this speed while reducing the Round-Trip Time (RTT) by \textbf{46.29\% compared to Cubic} and \textbf{32.40\% compared to New Reno}.
    \item \textbf{Reliability and Bufferbloat:} A critical advantage observed was the algorithm's stability. While TCP Cubic caused significant packet loss ($\approx 500$ packets) due to aggressive window growth and buffer overflow, the DRL agent maintained \textbf{near-zero packet loss}, effectively mitigating bufferbloat.
\end{itemize}

These findings confirm that Deep Reinforcement Learning can provide a robust solution for complex congestion control problems. The algorithm demonstrated superior adaptability, offering "high-quality" transmission (low latency, high reliability) rather than just "high-speed" transmission that sacrifices stability. This makes the proposed solution particularly significant for modern network infrastructures, such as 5G/6G mobile networks and IoT environments, where low latency is often more critical than raw bandwidth.

Although the results are promising, the performance of the RL algorithm depends on the training environment. Future research will focus on:
\begin{enumerate}
    \item Testing the agent in lossy wireless networks to further exploit its reliability advantage.
    \item Implementing more complex reward functions that explicitly penalize jitter.
    \item Evaluating the algorithm in large-scale topologies with competing TCP flows.
\end{enumerate}

In conclusion, this study provides compelling proof that DRL offers a powerful paradigm for developing highly adaptable TCP congestion control algorithms that prioritize network efficiency and user experience over aggressive bandwidth consumption.

\section{Recommendations}
Based on the findings, the following recommendations are proposed:

\begin{itemize}
    \item \textbf{Focus on Latency-Sensitive Applications:} Given the algorithm's success in reducing RTT by over 46\%, it should be tested in real-time applications such as cloud gaming, VoIP, and autonomous vehicle communication.
    \item \textbf{Exploration of "Chaos" Scenarios:} Since the agent showed high stability against buffer overflows, future tests should introduce random packet losses (simulating poor wireless signals) to verify if the DRL agent outperforms loss-based algorithms like Cubic in error-prone environments.
    \item \textbf{Hyperparameter Optimization:} Employing automated tuning (e.g., Bayesian optimization) for the reward function coefficients ($\alpha, \beta$) could further refine the throughput-delay balance.
    \item \textbf{Real-World Deployment:} Transitioning from NS-3 simulations to a Linux Kernel module implementation to test the algorithm on physical hardware testbeds.
\end{itemize}

\bibliographystyle{icstnum.bst}
\bibliography{references}

@article{subramanya2014performance,
author = {P, Subramanya and Ks, Vinayaka and H L, Gururaj and B, Ramesh},
year = {2014},
month = {01},
pages = {48-53},
title = {Performance Evaluation of High Speed TCP Variants in Dumbbell Network},
volume = {16},
journal = {IOSR Journal of Computer Engineering},
doi = {10.9790/0661-16264853}
}

@article{jacobson1988congestion,
author = {Jacobson, V.},
title = {Congestion avoidance and control},
year = {1988},
isbn = {0897912799},
publisher = {Association for Computing Machinery},
address = {New York, NY, USA},
url = {https://doi.org/10.1145/52324.52356},
doi = {10.1145/52324.52356},
pages = {314–329},
numpages = {16},
location = {Stanford, California, USA},
series = {SIGCOMM '88}
}

@article{cardwell2017bbr,
author = {Cardwell, Neal and Cheng, Yuchung and Gunn, C. and Yeganeh, Soheil and Jacobson, Van},
year = {2017},
month = {01},
pages = {58-66},
title = {BBR: Congestion-based congestion control},
volume = {60},
journal = {Communications of the ACM},
doi = {10.1145/3009824}
}

@article{GOMEZ2023110054,
title = {A survey on network simulators, emulators, and testbeds used for research and education},
journal = {Computer Networks},
volume = {237},
pages = {110054},
year = {2023},
issn = {1389-1286},
doi = {https://doi.org/10.1016/j.comnet.2023.110054},
url = {https://www.sciencedirect.com/science/article/pii/S1389128623004991},
author = {Jose Gomez and Elie F. Kfoury and Jorge Crichigno and Gautam Srivastava},
}

@inproceedings{Font2010,
author = {Font, Juan and Iñigo, Pablo and Domínguez-Morales, Manuel and Sevillano, Jose Luis and Amaya Rodriguez, Claudio},
year = {2010},
month = {04},
pages = {109},
title = {Architecture, design and source code comparison of ns-2 and ns-3 network simulators},
doi = {10.1145/1878537.1878651}
}

@article{toor2017,
author = {Toor, Amanjot and Jain, A.K.},
year = {2017},
month = {03},
pages = {62-69},
title = {A Survey on Wireless Network Simulators},
volume = {6},
journal = {Bulletin of Electrical Engineering and Informatics},
doi = {10.11591/eei.v6i1.568}
}

@article{zarrad2017,
author = {Zarrad, Anis and Alsmadi, Izzat},
year = {2017},
month = {10},
pages = {155014771773821},
title = {Evaluating network test scenarios for network simulators systems},
volume = {13},
journal = {International Journal of Distributed Sensor Networks},
doi = {10.1177/1550147717738216}
}

@inproceedings{tan2020optimized,
author = {Tan, Ziya and Karakose, Mehmet},
booktitle={2020 IEEE International Symposium on Systems Engineering (ISSE)},
year = {2020},
month = {10},
pages = {1-4},
title = {Optimized Deep Reinforcement Learning Approach for Dynamic System},
doi = {10.1109/ISSE49799.2020.9272245}
}

@article{littman2015reinforcement,
author = {Littman, Michael},
year = {2015},
month = {05},
pages = {445-51},
title = {Reinforcement learning improves behaviour from evaluative feedback},
volume = {521},
journal = {Nature},
doi = {10.1038/nature14540}
}

@article{puterman2000tutorial,
  title={A tutorial survey of reinforcement learning},
  author={Puterman, Martin L},
  journal={INFORMS Journal on Computing},
  volume={12},
  number={3},
  pages={187--209},
  year={2000},
  doi={10.1287/ijoc.12.3.187.16484}
}

@article{ee2007complex,
  title={Complex network analysis of communication networks},
  author={Ee, Maurice van and Heijden, Ger van der},
  journal={Physica A: Statistical Mechanics and its Applications},
  volume={379},
  number={1},
  pages={269--277},
  year={2007},
  doi={10.1016/j.physa.2007.03.011}
}

@inproceedings{Jiang2015/11,
  title={A review of Network Topology},
  author={RuoJing Jiang},
  year={2015/11},
  booktitle={Proceedings of the 2015 4th International Conference on Computer, Mechatronics, Control and Electronic Engineering},
  pages={1167-1170},
  issn={2352-5401},
  isbn={978-94-6252-110-0},
  url={https://doi.org/10.2991/iccmcee-15.2015.222},
  doi={10.2991/iccmcee-15.2015.222},
  publisher={Atlantis Press}
}

@book{ros2005protocole,
  title={Protocole de transport TCP},
  author={Ros, David},
  year={2005},
  publisher={Ed. Techniques Ing{\'e}nieur}
}

@article{pujolle2015architecture,
author = {Pujolle, Guy},
year = {2015},
month = {08},
pages = {},
title = {Architecture TCP/IP},
journal = {Réseaux Télécommunications},
doi = {10.51257/a-v1-h2288}
}

@book{murhammer1998tcp,
  title={TCP/IP tutorial and technical overview},
  author={Murhammer, Martin W and Atakan, Orcun and Bretz, Stefan and Pugh, Larry R and Suzuki, Kazunari and Wood, David H},
  year={1998},
  publisher={Prentice Hall Upper Saddle River, NJ}
}

@misc{handley2000rfc2861,
    series =    {Request for Comments},
    number =    2861,
    howpublished =  {RFC 2861},
    publisher = {RFC Editor},
    doi =       {10.17487/RFC2861},
    url =       {https://www.rfc-editor.org/info/rfc2861},
    author =    {Jitendra Padhye and Sally Floyd and Mark J. Handley},
    title =     {{TCP Congestion Window Validation}},
    pagetotal = 11,
    year =      2000,
    month =     jun,
}

@article{welzl2006network,
author = {Welzl, Michael},
year = {2006},
month = {05},
pages = {1-263},
title = {Network Congestion Control: Managing Internet Traffic},
isbn = {9780470025284},
journal = {Network Congestion Control: Managing Internet Traffic},
doi = {10.1002/047002531X}
}

@ARTICLE{Akpakwu10891783,
  author={Akpakwu, Godfrey A. and Mathonsi, Topside E. and Tshilongamulenzhe, Tshimangadzo M. and Maswikaneng, Solly P. and Muchenje, Tonderai},
  journal={IEEE Access}, 
  title={Congestion Control in Constrained Application Protocol for the Internet of Things: State-of-the-Art, Challenges, and Future Directions}, 
  year={2025},
  volume={13},
  number={},
  pages={33733-33767},
  doi={10.1109/ACCESS.2025.3543415}}

@INPROCEEDINGS{Kuzmanovic1209192,
  author={Kuzmanovic, A. and Knightly, E.W.},
  booktitle={IEEE INFOCOM 2003. Twenty-second Annual Joint Conference of the IEEE Computer and Communications Societies (IEEE Cat. No.03CH37428)}, 
  title={TCP-LP: a distributed algorithm for low priority data transfer}, 
  year={2003},
  volume={3},
  number={},
  pages={1691-1701 vol.3},
  doi={10.1109/INFCOM.2003.1209192}}

@article{lorincz2021comprehensive,
author = {Lorincz, Josip and Klarin, Zvonimir and Ozegovic, Julije},
year = {2021},
month = {06},
pages = {1-41},
title = {A Comprehensive Overview of TCP Congestion Control in 5G Networks: Research Challenges and Future Perspectives},
volume = {21},
journal = {Sensors},
doi = {10.3390/s21134510}
}

@article{abdullah2023improving,
author = {Abdullah, Saleh and Farag, Mohamed and Abd elkader, Hatem and Abo-Youssef, S.},
year = {2023},
month = {04},
pages = {228-235},
title = {Improving the TCP Newreno Congestion Avoidance Algorithm on 5G Networks},
journal = {Journal of Communications},
doi = {10.12720/jcm.18.4.228-235}
}

@inproceedings{kanagarathinam2018d,
author = {Kanagarathinam, Madhan and Singh, Sukhdeep and Irlanki, Sandeep and Roy, Abhishek and Saxena, Navrati},
year = {2018},
month = {01},
pages = {},
title = {D-TCP: Dynamic TCP Congestion Control Algorithm for Next Generation Mobile Networks},
organization={IEEE},
booktitle={2018 15th IEEE Annual Consumer Communications \& Networking Conference (CCNC)},
doi = {10.1109/CCNC.2018.8319185}
}

@article{poorzare2021fb,
author = {Poorzare, Reza and Calveras, Anna},
year = {2021},
month = {06},
pages = {82812 - 82832},
title = {FB-TCP: a 5G mmWave Friendly TCP for Urban Deployments},
volume = {9},
journal = {IEEE Access},
doi = {10.1109/ACCESS.2021.3087239}
}

@article{midhula2023adaptive,
author = {K S, Midhula and P, Arun},
year = {2023},
month = {01},
pages = {1-1},
title = {An Adaptive Congestion Control Protocol for Wireless Networks Using Deep Reinforcement Learning},
volume = {PP},
journal = {IEEE Transactions on Network and Service Management},
doi = {10.1109/TNSM.2023.3325543}
}

@article{bazmi2014neural,
author = {Bazmi, Parisa and Keshtgary, Manijeh},
year = {2014},
month = {10},
pages = {214},
title = {A neural network based congestion control algorithm for content-centric networks},
volume = {3},
journal = {Journal of Advanced Computer Science \& Technology},
doi = {10.14419/jacst.v3i2.3696}
}

@article{topalova2018control,
  title={Control of traffic congestion with weighted random early detection and neural network implementation},
  author={Topalova, Irina and Radoyska, Pavlinka},
  journal={ICAS 2018},
  volume={16},
  year={2018}
}

@article{van2015deep,
author = {Van Hasselt, Hado and Guez, Arthur and Silver, David},
year = {2015},
month = {09},
pages = {},
title = {Deep Reinforcement Learning with Double Q-Learning},
volume = {30},
journal = {Proceedings of the AAAI Conference on Artificial Intelligence},
doi = {10.1609/aaai.v30i1.10295}
}

@article{mnih2015human,
author = {Mnih, Volodymyr and Kavukcuoglu, Koray and Silver, David and Rusu, Andrei and Veness, Joel and Bellemare, Marc and Graves, Alex and Riedmiller, Martin and Fidjeland, Andreas and Ostrovski, Georg and Petersen, Stig and Beattie, Charles and Sadik, Amir and Antonoglou, Ioannis and King, Helen and Kumaran, Dharshan and Wierstra, Daan and Legg, Shane and Hassabis, Demis},
year = {2015},
month = {02},
pages = {529-33},
title = {Human-level control through deep reinforcement learning},
volume = {518},
journal = {Nature},
doi = {10.1038/nature14236}
}

@article{vidyasagar2023tutorial,
author = {Mathukumalli Vidyasagar},
title = {A tutorial introduction to reinforcement learning},
journal = {SICE Journal of Control, Measurement, and System Integration},
volume = {16},
number = {1},
pages = {172--191},
year = {2023},
publisher = {Taylor \& Francis},
doi = {10.1080/18824889.2023.2196033},
}

@Book{sutton1998reinforcement,
  author =	 {Richard S. Sutton and Andrew G. Barto},
  title =	 {Reinforcement Learning: An Introduction},
  publisher =	 {{MIT} Press},
  year =	 1998,
  url ={http://www.cs.ualberta.ca/\%7Esutton/book/ebook/the-book.html},
  isbn = 	 {0262193981},
  googleid = 	 {1Ubs6AcJ6QUJ:scholar.google.com/},
  cluster = 	 {425881569340442325},
}

@misc{floyd2004newreno,
  title        = {The NewReno Modification to TCP's Fast Recovery Algorithm},
  author       = {T. Henderson and S. Floyd and A. Gurtov and Y. Nishida},
  howpublished = {RFC 6582},
  year         = {2012},
  month        = apr,
  doi          = {10.17487/RFC6582},
  url          = {https://www.rfc-editor.org/info/rfc6582},
}

@book{kurose2022computer,
author = {Kurose, James F. and Ross, Keith W.},
address = {Harlow, United Kingdom},
booktitle = {Computer networking : a top-down approach},
edition = {Eighth edition.},
isbn = {9781292405469},
language = {eng},
publisher = {Pearson},
title = {Computer networking : a top-down approach },
year = {2022 - 2022},
}

@article{padhye1998modeling,
author = {Padhye, J. and Firoiu, V. and DF, Towsley and Kurose, Jim},
year = {2000},
month = {01},
pages = {},
title = {Modeling TCP through-put: A simple model and its empirical validation},
volume = {28},
journal = {Computer Communication Review}
}

@article{ha2008cubic,
author = {Ha, Sangtae and Rhee, Injong and Xu, Lisong},
year = {2008},
month = {07},
pages = {64-74},
title = {CUBIC: a new TCP-friendly high-speed TCP variant},
volume = {42},
journal = {Operating Systems Review},
doi = {10.1145/1400097.1400105}
}

@article{bruhn2023performance,
  title={Performance and improvements of TCP CUBIC in low-delay cellular networks},
  author={Bruhn, Philipp and others},
  journal={Computer Networks},
  volume={224},
  pages={109609},
  year={2023},
  doi={10.1016/j.comnet.2023.109609}
}

@inproceedings{afonin2019development,
author = {Afonin, Igor and Gorelik, A. and Muratchaev, Said and Volkov, A. and Morozov, E.},
year = {2019},
month = {07},
pages = {1-5},
title = {Development of an adaptive TCP algorithm based on machine learning in telecommunication networks},
doi = {10.1109/SYNCHROINFO.2019.8814023}
}

@ARTICLE{wei2021congestion,
  author={Wei, Wenting and Gu, Huaxi and Li, Baochun},
  journal={IEEE Network}, 
  title={Congestion Control: A Renaissance with Machine Learning}, 
  year={2021},
  volume={35},
  number={4},
  pages={262-269},
  doi={10.1109/MNET.011.2000603}}

@inproceedings{kong2018improving,
  author    = {Kong, Yiming and Zang, Hui and Ma, Xiaoli},
  title     = {Improving TCP Congestion Control with Machine Intelligence},
  booktitle = {Proceedings of the 2018 Workshop on Network Meets AI \& ML (NetAI'18)},
  year      = {2018},
  pages     = {60--66},
  location  = {Budapest, Hungary},
  doi       = {10.1145/3229543.3229550}
}

@article{li2018qtcp,
author = {Li, Wei and Zhou, Fan and Chowdhury, Kaushik and Meleis, Waleed},
year = {2018},
month = {05},
pages = {1-1},
title = {QTCP: Adaptive Congestion Control with Reinforcement Learning},
volume = {PP},
journal = {IEEE Transactions on Network Science and Engineering},
doi = {10.1109/TNSE.2018.2835758}
}

@inproceedings{sneha2020prediction,
  author = {Sneha, Y. and Vimitha and Vishwasini and Boloor, Shravan and Adesh, N.},
  year = {2020},
  month = {10},
  pages = {188-193},
  title = {Prediction of Network Congestion at Router using Machine learning Technique},
  booktitle = {2020 IEEE International Conference on Distributed Computing, VLSI, Electrical Circuits and Robotics (DISCOVER)},
  doi = {10.1109/DISCOVER50404.2020.9278028}
}

@article{molia6reinforcement,
author = {Molia, Hardik},
year = {2024},
month = {10},
pages = {607-616},
title = {Reinforcement Learning based Adaptive Congestion Control for TCP over Wireless Networks},
volume = {11},
journal = {International Journal of Computer Networks and Applications},
doi = {10.22247/ijcna/2024/39}
}

@article{xiao2019tcp,
author = {Xiao, Kefan and Mao, Shiwen and Tugnait, Jitendra},
year = {2019},
month = {01},
pages = {1-1},
title = {TCP-Drinc: Smart Congestion Control Based on Deep Reinforcement Learning},
volume = {PP},
journal = {IEEE Access},
doi = {10.1109/ACCESS.2019.2892046}
}

@inproceedings{kasi2021tcp,
author = {Kasi, Shahrukh Khan and Das, Saptarshi and Biswas, Subir},
year = {2021},
month = {01},
pages = {1507-1513},
title = {TCP Congestion Control with Multiagent Reinforcement and Transfer Learning},
doi = {10.1109/CCWC51732.2021.9376056}
}

@inproceedings{hassan2022tcp,
author = {Hassan, Ali and Heydari, Shahram},
year = {2021},
month = {02},
pages = {306-311},
title = {TCP Congestion Avoidance in Data Centres using Reinforcement Learning},
doi = {10.23919/ICACT51234.2021.9370861}
}

@article{verma2020iot,
author = {Verma, Lal Pratap and Kumar, Mahesh},
year = {2019},
month = {12},
pages = {100157},
title = {An IoT based Congestion Control Algorithm},
volume = {9},
journal = {Internet of Things},
doi = {10.1016/j.iot.2019.100157}
}

@inproceedings{hamrioui2017load,
author = {Hamrioui, Sofiane and Lorenz, Pascal},
year = {2017},
month = {12},
pages = {1-6},
title = {Load Balancing Algorithm for Efficient and Reliable IoT Communications within E-Health Environment},
doi = {10.1109/GLOCOM.2017.8254435}
}

@Article{lim2020improving,
AUTHOR = {Lim, Chansook},
TITLE = {Improving Congestion Control of TCP for Constrained IoT Networks},
JOURNAL = {Sensors},
VOLUME = {20},
YEAR = {2020},
NUMBER = {17},
ARTICLE-NUMBER = {4774},
URL = {https://www.mdpi.com/1424-8220/20/17/4774},
PubMedID = {32846962},
ISSN = {1424-8220},
DOI = {10.3390/s20174774}
}

@misc{rfc9006,
  title        = {TCP Usage Guidance in the Internet of Things (IoT)},
  author       = {Carles Gomez and Jon Crowcroft and Michael Scharf},
  howpublished = {RFC 9006},
  year         = {2021},
  month        = may,
  publisher    = {RFC Editor},
  doi          = {10.17487/RFC9006},
  url          = {https://www.rfc-editor.org/info/rfc9006}
}

@article{ohashi2005development,
  title={Development of secured medical network with TCP2 for telemedicine},
  author={Ohashi, Kumiko and Gomi, Yuichiro and Nogawa, Hiroki and Mizushima, Hiroshi and Tanaka, Hiroshi},
  journal={Studies in Health Technology and Informatics},
  volume={116},
  pages={397--402},
  year={2005},
  publisher={Amsterdam; Washington, DC: IOS Press, 1991-}
}

@article{ahmad2018end,
author = {Ahmad, Mudassar and Hussain, Majid and Abbas, Beenish and Aldabbas, Omar and Jamil, Uzma and Ashraf, Rehan and Asadi, Shahla},
year = {2018},
month = {02},
pages = {1-1},
title = {End-to-End Loss Based TCP Congestion Control Mechanism as a Secured Communication Technology for Smart Healthcare Enterprises},
volume = {6},
journal = {IEEE Access},
doi = {10.1109/ACCESS.2018.2802841}
}

@article{kumar2020design,
author = {Kumar, Rajiv and Pandit, Shweta and Sharma, Ashutosh},
year = {2021},
month = {01},
pages = {456-457},
title = {Design of Reliable, Secure and Intelligent Systems for Healthcare Applications},
volume = {14},
journal = {Recent Patents on Engineering},
doi = {10.2174/187221211403201130093110}
}

\end{document}